\documentclass[aps,preprint,showpacs,preprintnumbers,amsmath,amssymb]{revtex4}
\usepackage{graphicx,amsfonts}
\usepackage{epsfig}
\usepackage{dcolumn}
\usepackage{bm}
\date{\today}

\begin{document}

\title{Stabilization of the Electroweak Scale in 3-3-1 Models\\ \vspace{1.5cm} }
\author{Alex G. Dias$^1$, V. Pleitez$^2$}
\affiliation{\vspace{0.5cm}\\ $^1$Centro de Ci\^encias Naturais e Humanas, Universidade Federal do ABC,\\
R. Santa Ad\'elia 166, Santo Andr\'e - SP, 09210-170, Brazil
\vspace{0.4cm} \\
$^2$ Instituto  de F\'\i sica Te\'orica, UNESP - S\~ao Paulo State University, \\
Caixa Postal 70532-2, S\~ao Paulo - SP, 01140-070, Brazil
\vspace{1.0cm}
}

\begin{abstract}
One way of avoiding the destabilization of the electroweak
scale through a strong coupled regime naturally occurs in models with a
Landau-like pole at the TeV scale. Hence, the quadratic divergence
contributions to the scalar masses are not considered as a problem
anymore since a new nonperturbative dynamic emerges at the TeV scale.
This scale should be an intrinsic feature of the models and
there is no need to invoke any other sort of protection for the
electroweak scale. In some models based on the
$SU(3)_C\otimes SU(3)_W\otimes U(1)_{X}$ gauge symmetry, a nonperturbative dynamics
arise and it stabilizes the electroweak scale.
\end{abstract}
\maketitle

%

Understanding the Standard Model (SM) of strong and electroweak
interactions as an effective description of a more fundamental theory
implies that, at an  energy scale denoted by $\Lambda$, new physics must arise.
Without any additional hypotheses, it is natural that
$v_{W}<\Lambda\leq M_{Pl}$, where $v_W =(\sqrt{2}G_F/2)^{1/2}\simeq 246$
GeV is the electroweak scale and $M_{Pl}\sim 10^{19}$ GeV is the Planck scale,
which is linked to the gravitational interactions according to semiclassical
arguments. However, since there is no symmetry at the quantum level protecting
the tree-level SM Higgs particle squared mass, $M^2_H=-2\mu^2$, it receives
quantum corrections so that
$m^2_H\rightarrow {\bar{M}}^2_H= -2(\mu^2 + \delta\mu^{2}) = 2(\lambda+\delta\lambda) v_W^2$.
Here $\mu^2$ and $\lambda$ are the usual tree-level  parameters of the
quadratic and the quartic terms in the renormalizable scalar potential
at tree level, $V_0= \mu^2 H^\dagger H+\lambda(H^\dagger H)^2$,
with $H$ the usual scalar doublet of the SM;
$\delta\mu^2$ and $\delta\lambda$ stand for the corresponding quantum
corrections to the tree level parameters. It so happens that $\delta\mu^2$ is proportional to
$\Lambda^2$, and $\delta\lambda$ is proportional to $\ln \Lambda$ \cite{meissner-plb07}.
Hence, the electroweak scale does not seem to be stable against quantum corrections unless some mechanism protects it. In the context of the SM, the Higgs,
$Z^0$ and $W$, bosons and the top quark give the dominant contributions
for the one-loop effective potential. Taking into account that $\Lambda$ is the cutoff appearing in the momentum integrals we have, with $g$, $g^{\prime}$, and $y_{T}$ the $SU(2)$, $U(1)_Y$, and the top quark coupling constants of the SM, respectively,
\begin{eqnarray}
\delta\mu^{2}&\approx& 3\left[2M_{W}^{2}+{M_{Z}^{2}+ M_{H}^{2}
-4M_{T}^{2}} \right]\left( \frac{\Lambda}{4\pi v_{W}}\right)^2\nonumber\\
&=& 3\left[\frac{3}{4}g^{2}+\frac{1}{4}g^{\prime 2}+2\lambda
-2y_{T}^{2} \right]\left( \frac{\Lambda}{4\pi }\right)^2,
 \label{e1}
\end{eqnarray}
plus terms proportional to $\ln \Lambda$ times loop factors.
Independently of the value of $\Lambda$, a Higgs boson with a tree level mass $M_H\approx 310$
GeV (i.e. $\lambda\approx 0.8$) could make $\delta\mu^{2}$ in Eq.~(\ref{e1}) near zero~\cite{veltman81}.
However, according to the electroweak precision data such a heavy Higgs seems not to be favored by the global fit of the SM~\cite{pdg,lep05}.

Ideas for stabilizing the electroweak scale
have been put forward in the past. The as\-sump\-tion of supersymmetry (SUSY) at the
TeV scale was probably the first of such ideas~\cite{kaul}. More recently,
solutions such as the Little Higgs (LH)~\cite{lh,lhreviews}
and the extra dimension (ED) models~\cite{ed} were suggested.
The LH models are constructions where the SM is contained
in the low energy spectrum with the Higgs boson as a pseudo-Nambu-Goldstone
boson of some particular global symmetry. This global symmetry protects
one-loop quadratic divergences in $\delta\mu^2$. In LH models the
common point resides on the fact that they are nonrenormalizable, defining,
in principle, an energy cutoff which is around
$\Lambda=10$ TeV for the internal momentum integrals. It is interesting
that this mechanism can be implemented if global and local $SU(3)$
symmetries are considered. In fact, Ref.~\cite{lhiggs}
worked out a LH model which includes right-handed neutrinos
transforming nontrivially under $SU(3)_W\otimes U(1)_X$ gauge symmetry.
This was considered previously as an extension of the electroweak interactions in Ref.~\cite{mpp}.

Notwithstanding, there are other motivations for considering
$SU(3)_W\otimes U(1)_X$ symmetry as being realized at the TeV scale,
as in the so-called 3-3-1 models~\cite{mpp,331,331pt}. One important
point is that in some of these models~\cite{331,331pt}, depending
on the representation content, there is an intrinsic cutoff:
the $U(1)_X$ coupling constant gets a Landau-like pole at energies of the order
of few TeV~\cite{phf,dasjain,pl331,evpl331}. Thus, the theory becomes strongly
coupled at the TeV scale inducing, in this way, a natural cutoff for
the quantum corrections.

One of the reasons that $\Lambda$ can be arbitrarily high in
the SM is that none of its gauge coupling constants reaches a value
which invalidates the perturbative  expansion at a testable energy.
Let us imagine that a new particle content, which arises from a symmetry
struc\-tu\-re already revealed at the TeV scale, drives
the gauge coupling constants towards a region of values where the theory is
nonperturbative. This would indicate that the theory goes to a strongly coupled
regime for higher energies.
That is the case if the electroweak sector of the SM is embedded into an
$SU(3)_W\otimes U(1)_{X}$ symmetry, as in a specific class of 3-3-1 models
~\cite{331,331pt}.  Their symmetry reduction, $SU(3)_W\otimes U(1)_{X}$
$\rightarrow$ $SU(2)_W\otimes U(1)_{Y}$, to the SM group is realized through a
scalar field condensation by a vacuum expectation value $\langle \chi^0 \rangle$ $=v_\chi$,
related to the energy scale denoted $\mu_{331}$ according to
$\sqrt2 \,v_\chi\approx \mu_{331}$. Below $\mu_{331}$ we have
an effective $SU(2)_W\otimes U(1)_{Y}$ symmetry with the SM
fermions, gauge boson multiplets, and two scalar doublets composing the light
degrees of freedom. These are the active degrees of freedom below $\mu_{331}$.
All other fields are presumed to be heavy, i. e., with masses around $\mu_{331}$.

In these 3-3-1 models there is the relation
\begin{equation}
\alpha_{_X}(\mu_{331}) = \frac{\alpha(\mu_{331})}{1-4\sin^2\theta_W(\mu_{331})},
\label{polo}
\end{equation}
where $\alpha_{_X}=g^2_{_X}/4\pi$ is the gauge coupling constant of the $U(1)_X$
gauge factor, written in terms of the electroweak mixing angle  $\theta_W$, and the electromagnetic coupling $\alpha$, both defined at the $\mu_{331}$ scale.
This relation is used to determine the initial value of $\alpha_{_X}(\mu_{331})$, making the evolution of $\sin^2\theta_W\equiv\sin^2\theta_W(M_Z)\approx 0.231$ and
$\alpha\equiv\alpha(M_Z)\approx 1/128$ from the $Z^0$ pole to the $\mu_{331}$ scale.
We see that a Landau-like pole will be developed as $\sin^2\theta_W(\mu)$
evolves to the value 0.25. This value is reached for $\mu\approx 4.2$ TeV, making the evolution only with the active degrees of freedom below $\mu_{331}$. This means that even if $\mu_{331}$ is above this value the cutoff in Eq.~(\ref{e1}) must be such that $\Lambda < 4.2$ TeV. Before reaching this pole, $\alpha_{_X}$ goes outside the perturbative regime, and we cannot draw any conclusion based on perturbation theory. From the evolution equation for
$\alpha_X$, with the initial point as in Eq.~(\ref{polo}), the upper energy
limit $\bar{\Lambda}<\Lambda$ were perturbative treatment loses its validity, i. e.,
$\alpha_X(\bar{\Lambda})\approx 1$, is
\begin{equation}
\bar{\Lambda}=\mu_{_{331}}\left(\frac{M_{Z}}{\mu_{_{331}}}
\right)^{\frac{7}{13}}\,e^{\frac{\pi}{13}
\left[\frac{1}{\alpha}[1-4\sin^{2}\theta_{W}]
-1\right]}.
\label{ptlim}
\end{equation}

The model predicts a mass relation between the neutral $Z^\prime$ and the double charged $M_{U}$ gauge bosons,
\begin{equation}
\frac{M^2_{Z^\prime}}{M^2_{U}}\approx \frac{4\cos^{2}\theta_{W}}{3-12\sin^{2}\theta_{W}},
\label{massrel}
\end{equation}
with $M^2_{U}\approx g^2\mu_{331}^2/8$. So, using the lower bound for the $Z^\prime$ mass obtained in Ref.~\cite{boundzlya2008} as
$M_{Z^\prime}\geq 620 $ GeV, which implies the minimal value
$\mu_{_{331}}^{min}\approx 750$ GeV in Eq.~(\ref{ptlim}), we have
$\bar{\Lambda}\approx 2$ TeV. On the other hand, using the lower
bound of $M_{U}\geq 750$ GeV~\cite{tully} for the mass of the
double charged gauge boson in the model, $\mu_{_{331}}^{min}\approx 3.2$
TeV so that $\bar{\Lambda}=4$ TeV. It was pointed out in Ref.~\cite{bilepya2009} that
even for $M_{Z^\prime}\simeq 1.4 $, thousands of new single charged
vector bosons presented in the model could be produced at the LHC,
pointing out the possibility of distinguishing this model from other
models. For other aspects of gauge boson phenomenology, see also~\cite{pheno-gb}.

Here we will use the effective potential~\cite{cw,sher89} in the formalism of Ref.~\cite{jackiw74}.
Since $\Lambda$ is an upper limit for evaluating the integrals,  omitting constant terms
proportional to $\Lambda^4$,  the one-loop contribution to the effective potential is
{\small{
\begin{eqnarray}
V_1(h) &\approx & \frac{1}{64\pi^2 }\sum_i n_{i}\left [2\Lambda^2 {\bf Tr}
\,({\bf M^\dagger_i M_i}) + {\bf Tr}\left\{ ({\bf M^\dagger_i
M_i})^2\left (\ln  \frac{{\bf M^\dagger_i M_i}}{\Lambda^2}
 -\frac{1}{2}\right) \right\}\right ],
\label{ep}
\end{eqnarray}
}}
where $n_i$ is the number of degrees of freedom of the field $i$,
including a minus sign for fermions; ${\bf M_i \equiv M_i}(h)$ are
obtained from the tree-level mass matrices of the model using
$v_W=h$. We will consider below only  terms proportional to $\Lambda^2$, since
they are the most relevant contributions for our purposes.

The model we consider has an approximate global $SU(3)_L \otimes SU(3)_R$ symmetry~\cite{newp}.
In the scalar sector this global symmetry is supposed to be exact.
Defining the tritriplet $\Phi=(\eta\;\rho\;\chi)$ transforming as
$\Phi\rightarrow\Omega_L\Phi\,\Omega^\dagger_R$ under $SU(3)_L \otimes SU(3)_R$,
the scalar potential is
\begin{eqnarray}
V[\Phi] &=& \mu^2\textrm{Tr}(\Phi^\dagger\Phi)+
{\lambda}_1[\textrm{Tr}(\Phi^\dagger\Phi)]^2+
{\lambda}_2\textrm{Tr}(\Phi^\dagger\Phi)^2 +\frac{f}{6}\epsilon_{ijk}\epsilon_{mnl}(\Phi_{im} \Phi_{jn}\Phi_{kl}+H.c.).
\label{potlr}
\end{eqnarray}

The Yukawa interactions and the gauge interactions introduced by $U(1)_X$ symmetry
explicitly break $SU(3)_L \otimes SU(3)_R$. For the
Yukawa Lagrangian we have
\begin{eqnarray}
-\mathcal{L}_Y&=&\bar{Q}_{iL}[g^u_{i\alpha} u_{\alpha R}\rho^*+
g^d_{i\alpha} d_{\alpha R}\eta^*+g^j_{ik}j_{kR}\chi^*]+\bar{Q}_{3L}[y^u_\alpha u_{\alpha R}\eta+y^d_\alpha d_{\alpha R}\rho+y^JJ_R\chi]
\nonumber \\
&+&\bar{\Psi}_{aL}[ g^\nu_{ab}\nu_{bR}\eta^*+
g^l_{ab}l_{bR}\rho^*+g^E_{ab}E_{bR}\chi^*]+H.c.,
\label{yuka331}
\end{eqnarray}
where repeated indices are to be summed accordingly with the fields. Terms
like $y_{ab}\epsilon_{ijk}{\overline{\Psi}}_{iaL}({\Psi_j}_{bL})^c\eta_k$ and
$m_{ab}\nu_{aR}(\nu_{bR})^c$ are not relevant for us here. We assume these terms
are forbidden by some symmetry. The global and local symmetries are spontaneously
broken with the vacuum expectation value for $\Phi$
\begin{equation}
\langle\Phi\rangle=\frac{1}{\sqrt2}\,
diag\left( \begin{array}{ccc}
v_\eta, & v_\rho,  & v_\chi \\
\end{array}\right),
\label{c6}
\end{equation}
leaving only the electromagnetic $U(1)$ factor as local symmetry.

In the Appendix the tree-level ${\bf M_i}(v_\eta,v_\rho,v_\chi)$ matrices
needed to obtain the quadratic corrections for the present model are shown.
Taking the first term in Eq.~(\ref{ep}) and the trace of the matrices in the
Appendix, we get the one-loop corrections for the bilinears in the potential

\begin{equation}
V_{_{eff}}=(\mu^2+\delta\mu^2_\eta)\eta^\dagger \eta+
(\mu^2+\delta\mu^2_\rho)\rho^\dagger \rho+
(\mu^2+\delta\mu^2_\chi)\chi^\dagger \chi+...
\label{vb}
\end{equation}
with
\small{
\begin{eqnarray}
&  \delta\mu^2_\eta &\approx \left[ 4g^2+20\lambda_1+12\lambda_2
-6\sum_{\alpha=1}^3(\vert y^{u}_{\alpha}\vert^2
+ \vert g^{d}_{1\alpha}\vert^2 + \vert g^{d}_{2\alpha}\vert^2)\right]
\left( \frac{\Lambda}{4\pi }\right)^2, \nonumber\\ \nonumber\\
& \delta\mu^2_\rho &\approx \left[4g^2+3g^2_{_X} +20\lambda_1+
12\lambda_2-6\sum_{\alpha=1}^3(\vert g^{u}_{1\alpha}\vert^2
+ \vert g^{u}_{2\alpha}\vert^2+ \vert y^{d}_{\alpha}\vert^2)\right]
\left( \frac{\Lambda}{4\pi }\right)^2, \nonumber\\ \nonumber\\
& \delta\mu^2_\chi &\approx \left[ 4g^2+3g^2_{_X}+20\lambda_1+
12\lambda_2-2\sum_{a=1}^3\vert g^{E}_{aa}\vert^2-6\vert y^{J}\vert^2 -6
\sum_{i=1}^2\vert g^{j}_{i\,i}\vert^2
\right]
\left( \frac{\Lambda}{4\pi }\right)^2.\nonumber\\
\label{dmuss}
\end{eqnarray}
}
Observe that the global $SU(3)_L \otimes SU(3)_R$ is recovered when $g_{_X}$
and the Yukawa couplings in Eq.~(\ref{yuka331}) are made equal to zero
resulting in $\delta\mu^2_\eta$ = $\delta\mu^2_\rho$ = $\delta\mu^2_\chi$,
as it should be.

For the potential in Eq.~(\ref{potlr}) we have to have $v_\eta=v_\rho=v_\chi$.
The quantum corrections bring into the constraint equations, Eqs.~(\ref{vinc}) below,
the effects of th explicit breakdown of $SU(3)_L \otimes SU(3)_R$, making it possible to
have $v_\eta\not=v_\rho\not=v_\chi$. Considering  the dominant contributions in
Eqs.~(\ref{dmuss}) and disregarding the corrections for the couplings in the
self-interaction terms, the constraint
equations for minimizing the potential are
\begin{eqnarray}
& & -(\mu^2+\delta\mu^2_\eta) = \lambda_2v_\eta^2 + \lambda_1v^2
+ \frac{\sqrt2}{4}\frac{f}{v_\eta}v_\rho v_\chi,\nonumber\\
& & -(\mu^2+\delta\mu^2_\rho)=\lambda_2v_\rho^2 + \lambda_1v^2
+ \frac{\sqrt2}{4}\frac{f}{v_\rho}v_\eta v_\chi, \nonumber\\
& & -(\mu^2+\delta\mu^2_\chi)=\lambda_2v_\chi^2 + \lambda_1v^2
+ \frac{\sqrt2}{4}\frac{f}{v_\chi}v_\rho v_\eta.
\label{vinc}
\end{eqnarray}
where $v^2=v_\eta^2+v_\rho^2+v_\chi^2$. Since $\delta\mu^2_\eta\not=\delta\mu^2_\rho\not=\delta\mu^2_\chi$ it is
possible to have $v_\eta\not=v_\rho\not=v_\chi$. Once the scale
$\Lambda$ is really limited in the model, we do not expect that any severe
fine-tuning is needed in Eqs. (\ref{vinc}).

Next we show the masses, at leading order, for some of the
scalar fields. There are two single charged scalars with masses given by
\begin{eqnarray}
M^2_{1+} = \left(\lambda_2-\frac{\sqrt2}{4}\frac{fv_\chi}{v_\eta v_\rho}\right)
(v_\eta^2+ v_\rho^2 ),\quad
M^2_{2+} = \left(\lambda_2-\frac{\sqrt2}{4}\frac{fv_\rho}{v_\eta v_\chi}\right)
(v_\eta^2+ v_\chi^2 );
\label{massch}
\end{eqnarray}
a double charged scalar with mass given by
\begin{eqnarray}
 M^2_{++} = \left(\lambda_2-\frac{\sqrt2}{4}\frac{fv_\eta}{v_\chi v_\rho}\right)
 (v_\rho^2 + v_\chi^2 );
\label{massdch}
\end{eqnarray}
and a pseudoscalar (CP odd) with mass given by
\begin{eqnarray}
 M^2_A = -\frac{\sqrt2}{4}f v_\eta v_\rho v_\chi\left(\frac{ 1}{v^2_\chi}+
 \frac{1}{v^2_\rho}+\frac{1}{v^2_\eta}\right).
\label{massps}
\end{eqnarray}
Assuming $v_\eta$, $v_\rho$, and $v_\chi$ are real and positive, we have that
$f <0$. This condition, along with $\lambda_2>0$, guarantees positive squared
masses also for the charged scalars, as we see from Eqs.~(\ref{massch}),
(\ref{massdch}). For the three CP even scalars, we
have not displayed their expression once they do not have simple closed form.

The existence of an ultraviolet singularity in one of the running coupling
constants, through a Landau-like pole, may indicate an energy scale at which
new phenomena could intervene. It is not clear at all what new phenomena
would arise at energies near or above the Landau-like
pole in these 3-3-1 models, but it could modify the running of the low energy
coupling constant. One possibility is the appearance of new particles from fields
forming representations affecting the running of the coupling constants such that the pole is avoided at reachable energies~\cite{evpl331}. An investigation with this hypothesis is needed
in order to see how the electroweak scale would then be stabilized in this case.

Let us compare this sort of 3-3-1 model with the SUSY, LH, and ED solutions for the
stabilization of the electroweak scale. SUSY, at the electroweak scale, is a \textit{renormalizable} theory; however, it is needed to assume that the scale related with SUSY and the masses of the supersymmetric partners be at the TeV scale. It is also useful for the unification of three of the fundamental forces, but it has trouble with the stabilization of the proton~\cite{murayama2002}.
The LH models are nonlinear realization of the spontaneous symmetry
breaking of a global symmetry, thus they are \textit{nonrenormalizable} theories
and remain perturbative, by construction, until an energy scale of the order of 10 TeV;
to have naturalness beyond this scale, some similar mechanism has to be invoked, i.e.,
a second LH model, and so on.  The ED proposals are also \textit{nonrenormalizable} theories, and the energy scale to solve the hierarchy problem is chosen by hand as well. The
3-3-1 models are different in the following sense: they are \textit{renormalizable} theories and
the energy scale at the TeV scale is an intrinsic property of the theories. We did not assumed that they have an appropriate value for solving the problem of the electroweak scale. The LH and ED solutions need ultraviolet completion; the 3-3-1 models do not. LH and ED solutions to the stabilization of the electroweak scale are \textit{ad hoc} since they are proposed \textit{just} to solve the problem. The 3-3-1 models were proposed for other reasons and have interesting consequences and predictive power~\cite{newp}.

Finally, we stress that the argument that the running of $\sin^{2}\theta_{W}$
gives an energy bound defining the model symmetry structure, as in
Eq.~(\ref{polo}), must be seen as a real prediction of this minimal
3-3-1 Model. It is not possible to arbitrarily rise the scale related
to the new particles in this model, which become inconsistent for energies
$\mu$ at the Standard Model level such that $\sin^{2}\theta_{W}(\mu)>0.25$.
This explains why $\sin^{2}\theta_{W}(M_Z)<0.25$ and also why
there are no threatening divergences for the mass of the Higgs boson.

We have discussed the stabilization of the electroweak scale in a specific 3-3-1 model \cite{331}.
The same arguments we present here are valid for the version of the model in
Ref.~\cite{331pt} which does not have heavy charged leptons, and a scalar sextet is mandatory for generating mass to the known charged lepton. It may be that there is a whole class of models where the arguments we have put forward here are valid.

A.G.D. thanks FAPESP for financial support. Both authors are
supported partially by CNPq under the processes 302045/2007-4 (A.G.D),
2007/04825-3, and 302102/2008-6 (V.P).

\appendix

\section{Mass matrices}
\label{a1}

{\it Scalar fields:}

Defining $X=(1/\sqrt2)(v_x+X_{_R}+iX_{_I})$, where $X=\eta,\rho,\chi$, we have for the mass
matrix of the neutral real scalars, in the basis $(\eta_{_R}\,\rho_{_R}\,\chi_{_R})$,

\begin{eqnarray}
{\bf M}^2_{_R}  &=&  \mu^2{\bf 1} \nonumber\\
&  +&{ \left[
\begin{array}{ccc}
 3\lambda_1v_\eta^2+ {\lambda_2}(2v_\eta^2+v^2)& 2\lambda_1v_\eta v_\rho +
 \frac{\sqrt2}{4}fv_\chi & 2 \lambda_1v_\eta v_\chi +\frac{\sqrt2}{4}fv_\rho \\
 2 \lambda_1v_\eta v_\rho +\frac{\sqrt2}{4}fv_\chi & 3\lambda_1v_\rho^2+
 {\lambda_2}(2v_\rho^2+v^2) & 2\lambda_1v_\rho v_\chi +\frac{\sqrt2}{4}fv_\eta \\
 2\lambda_1v_\eta v_\chi +\frac{\sqrt2}{4}fv_\rho  & 2\lambda_1v_\rho v_\chi +
 \frac{\sqrt2}{4}fv_\eta & 3\lambda_1v_\chi^2+ {\lambda_2}(2v_\chi^2+v^2)
\end{array}
\right]}\nonumber\\
\label{mr}
\end{eqnarray}

for the neutral pseudoscalars, in the basis $(\eta_{_I}\,\rho_{_I}\,\chi_{_I})$,

\begin{eqnarray}
 {\bf M}^2_{_I}  = \mu^2{\bf 1} +
 \left[
\begin{array}{ccc}
 \lambda_1v^2+\lambda_2v_\eta^2 & -\frac{\sqrt2}{4}fv_\chi & -\frac{\sqrt2}{4}fv_\rho \\
 -\frac{\sqrt2}{4}fv_\chi & \mu^2+\lambda_1v^2+\lambda_2v_\rho^2 & -\frac{\sqrt2}{4}fv_\eta\\
 -\frac{\sqrt2}{4}fv_\rho  & -\frac{\sqrt2}{4}fv_\eta & \lambda_1v^2+\lambda_2v_\chi^2
\end{array}
\right];
\label{mps}
\end{eqnarray}

for the single charged scalars, in the basis $(\eta^-_1\,\rho^-)$,

\begin{eqnarray}
 {\bf M}^2_{_{1+}}  =\mu^2{\bf 1} +
 \left[
\begin{array}{cc}
 \lambda_1v^2+\lambda_2(v^2-v_\chi^2) &  \lambda_2v_\eta v_\rho -\frac{\sqrt2}{4}fv_\chi \\
 \lambda_2v_\eta v_\rho -\frac{\sqrt2}{4}fv_\chi  & \lambda_1v^2+\lambda_2(v^2-v_\chi^2)
\end{array}
\right],
\label{mch1}
\end{eqnarray}
and, in the $(\eta^-_2\,\chi^-)$ basis,
\begin{eqnarray}
 {\bf M}^2_{_{2+}}  =\mu^2{\bf 1} +
 \left[
\begin{array}{cc}
 \lambda_1v^2+\lambda_2(v^2-v_\rho^2) &  \lambda_2v_\eta v_\chi -\frac{\sqrt2}{4}fv_\rho \\
 \lambda_2v_\eta v_\chi -\frac{\sqrt2}{4}fv_\rho  & \lambda_1v^2+\lambda_2(v^2-v_\rho^2)
\end{array}
\right];
\label{mch2}
\end{eqnarray}

and for the double charged scalars, in the basis $(\rho^{--}\,\chi^{--})$,

\begin{eqnarray}
 {\bf M}^2_{++}  =\mu^2{\bf 1} +
 \left[
\begin{array}{cc}
 \lambda_1v^2+\lambda_2(v^2-v_\eta^2) &  \lambda_2v_\rho v_\chi -\frac{\sqrt2}{4}fv_\eta \\
 \lambda_2v_\rho v_\chi -\frac{\sqrt2}{4}fv_\eta  & \lambda_1v^2+\lambda_2(v^2-v_\eta^2)
\end{array}
\right].
\label{mdch}
\end{eqnarray}

{\it Gauge bosons:}

For the real neutral gauge bosons in the
$(W^3_\mu $, $W^8_\mu $,  $B_\mu)$ basis the mass matrix is
\begin{eqnarray}
M^2_{n.g.b.}=\left(\begin{array}{ccc}
\frac{g^2}{4}({v}_\eta^2+{v}_\rho^2)\quad & \frac{\sqrt 3g^2}{12}(
{v}_\eta^2-{v}_\rho^2 ) & -\frac{g_{_X}g}{2} {v}_\rho^2  \\
\frac{\sqrt 3g^2}{12}({v}_\eta^2-{v}_\rho^2 )  & \frac{g^2}{12}({v}^2 + 3v_\chi^2)&
 \frac{\sqrt 3}{6}g_{_X}g({v}_\rho^2 +2v_\chi^2) \\
\frac{g^2}{12}({v}^2 + 3v_\chi^2)  &\quad\frac{\sqrt 3}{6}g_{_X}g({v}_\rho^2 +2v_\chi^2)
& g_{_X}^2({v}_\rho^2 +v_\chi^2)\end{array}\right);
\label{mnt}
\end{eqnarray}
and for the non-Hermitian gauge bosons
\begin{eqnarray}
M^2_W = \frac{g^2}{4}({v}_\eta^2 + v_\rho^2),\;\;
M^2_V = \frac{g^2}{4}({v}_\eta^2 + v_\chi^2), \;\;
M^2_U = \frac{g^2}{4}({v}_\rho^2 + v_\chi^2).
\label{mch}
\end{eqnarray}

{\it Fermion fields:}

For type $u$ and $d$ quarks
\begin{eqnarray}
 M_u=
\frac{1}{\sqrt2}\left(\begin{array}{ccc}
g^u_{11}v^*_\rho & g^u_{12}v^*_\rho & g^u_{13}v^*_\rho\\
g^u_{21}v^*_\rho & g^u_{22}v^*_\rho & g^u_{23}v^*_\rho\\
y^u_{1}v_\eta & y^u_{2}v_\eta & y^u_{3}v_\eta
\end{array}
\right),
\,\, M_d=\frac{1}{\sqrt2}\left(\begin{array}{ccc}
g^d_{11}v^*_\eta & g^d_{12}v^*_\eta & g^d_{13}v^*_\eta\\
g^d_{21}v^*_\eta & g^d_{22}v^*_\eta & g^d_{23}v^*_\eta\\
y^d_{1}v_\rho & y^d_{2}v_\rho & y^d_{3}v_\rho
\end{array}
\right);
\label{massas}
\end{eqnarray}
for the $j_i$ and $J$ quarks, and exotic leptons
\begin{equation}
M_j=\left(\begin{array}{cc}
g^j_{11}& g^j_{12}\\
g^j_{21}& g^j_{22}
\end{array}\right)\frac{v_\chi}{\sqrt2}, \quad\quad M_J=\frac{y^J}{\sqrt2}v_\chi;\quad
M_E=\left(\begin{array}{ccc}
g^E_{11} & g^E_{12} & g^E_{13}\\
g^E_{21} & g^E_{22} & g^E_{23}\\
g^E_{31} & g^E_{32} & g^E_{33}
\end{array}
\right)\frac{v_\chi}{\sqrt2};
 \label{mj}
\end{equation}

\end{document}